\documentclass[12pt,a4paper]{article}

\usepackage{cite}
\usepackage{graphicx}
\usepackage{epsfig}
\usepackage{amssymb}
\usepackage{amsmath}
\usepackage{latexsym}

\newcommand{\capdef}{}
\newcommand{\mycaption}[2][\capdef]{\renewcommand{\capdef}{#2}%
       \caption[#1]{{\footnotesize #2}}}
\makeatletter
\renewcommand{\fnum@table}{\textbf{\tablename~\thetable}}
\renewcommand{\fnum@figure}{\textbf{\figurename~\thefigure}}
\makeatother

\makeatletter
\renewcommand{\section}{\@startsection{section}{1}{0em}{-\baselineskip}%
{\baselineskip}{\normalfont\large\bfseries}}
\renewcommand{\subsection}%
{\@startsection{subsection}{2}{0em}{-0.7\baselineskip}%
{0.7\baselineskip}{\normalfont\bfseries}}
\makeatother

\renewcommand{\baselinestretch}{1.07}

\newcommand{\stheta}{\ensuremath{\sin^22\theta_{13}}}

\newcommand{\delCP}{\ensuremath{\delta_{\rm CP}}}
\newcommand{\alphat}{\tilde\alpha}
\newcommand{\He}{\ensuremath{^6{\mathrm{He}}}}
\newcommand{\Ne}{\ensuremath{^{18}{\mathrm{Ne}}}}

\begin{document}
%%%%%%%%%%%%%%%%%%%%%%%%%%%%%%%%%%%%%%%%%%%%%%%%%%%%%%%%%%%%%%%%%%%%%
%%%%                     Title-page                              %%%%
%%%%%%%%%%%%%%%%%%%%%%%%%%%%%%%%%%%%%%%%%%%%%%%%%%%%%%%%%%%%%%%%%%%%%

%\begin{titlepage}

% the footnote symbols are only redefined for the title page !
\renewcommand{\thefootnote}{\alph{footnote}}

\begin{flushright}
CERN-PH-TH/2007-060\\
\end{flushright}

\vspace*{0.5cm}

\renewcommand{\thefootnote}{\fnsymbol{footnote}}
\setcounter{footnote}{0}

\begin{center} 
\Large\textbf{
Determination of the neutrino mass hierarchy in the regime of small matter effect
}
\end{center}

\vspace*{0.5cm}

\begin{center} 
{\bf Thomas Schwetz\footnote{email: \tt schwetz AT cern.ch}}

%\vspace*{2mm}
\it CERN, Physics Department, Theory Division, CH-1211 Geneva 23, Switzerland
\end{center}

\vspace*{0.3cm}

\begin{abstract}
We point out a synergy between T-conjugated oscillation channels in
the determination of the neutrino mass hierarchy with oscillation
experiments with relatively short baselines ($L \lesssim 700$~km),
where the matter effect is small. If information from all four
oscillation channels $\nu_\mu\to\nu_e$, $\bar\nu_\mu\to\bar\nu_e$,
$\nu_e\to\nu_\mu$ and $\bar\nu_e\to\bar\nu_\mu$ is available, a matter
effect of few percent suffices to break the sign-degeneracy and allows
to determine the neutrino mass hierarchy. The effect is discussed by
analytical considerations of the relevant oscillation probabilities,
and illustrated with numerical simulations of realistic experimental
setups.
Possible configurations where this method could be applied are the
combination of a super beam experiment with a beta beam or a neutrino
factory, or a (low energy) neutrino factory using a detector with
muon and electron charge identification.
%\pacs{14.60.Pq, 14.60.Lm}
\end{abstract}

\renewcommand{\thefootnote}{\arabic{footnote}}
\setcounter{footnote}{0}

%\newpage
%\tableofcontents

\vspace*{0.5cm}

%%%%%%%%%%%%%%%%%%%%%%%%%%%%%%%%
\section{Introduction}
%%%%%%%%%%%%%%%%%%%%%%%%%%%%%%%%

The determination of the neutrino mass hierarchy is one of the most
interesting open issues in neutrino physics. Present data allow for
the two possibilities normal hierarchy (NH) and inverted hierarchy
(IH), which are conventionally parametrized by the sign of the
difference of the mass-squares of the first and third neutrino mass
eigenstates: $\Delta m^2_{31} > 0$ for NH and $\Delta m^2_{31} < 0$
for IH. Identifying which of these two possibilities is realized in
nature is of great importance for our understanding of the neutrino
mass mechanism, the relation of neutrinos to the charged fermions, and
the problem of flavour in general.

Nevertheless, the determination of the sign of $\Delta m^2_{31}$ turns
out to be experimentally challenging. The most promising way seems to
be to explore the matter effect in neutrino
oscillations~\cite{Wolfenstein:1977ue,Barger:1980tf,Mikheev:1986gs}. This
can be done with long-baseline experiments~\cite{Freund:1999gy},
atmospheric neutrinos~\cite{atm}, or supernova
neutrinos~\cite{supernova}. Alternative methods to determine the
hierarchy, not based on the matter effect, have been
proposed~\cite{non-matter,DBD}, but they turn out to be extremely
challenging, if not impossible in practice.

The usual strategy to determine the mass hierarchy in long-baseline
experiments is to consider configurations where the matter effect is
{\it strong}, with baselines as long as possible, ideally several
1000~km (see, e.g., ref.~\cite{Gandhi:2004bj} for a recent work), and
neutrino energies close to (or above) the resonance
energy~\cite{Mikheev:1986gs}, which for typical densities of earth
matter and $|\Delta m^2_{31}| \simeq 2.5\times 10^{-3}$~eV$^2$ is
around 10~GeV. The basic idea is to observe the effect of the
resonance, which occurs for neutrinos in case of NH and anti-neutrinos
for IH. Hence, the discrimination of the two hierarchies is based on
the difference of the matter effect in CP-conjugated oscillation
channels, e.g., $\nu_e\to\nu_\mu$ and $\bar\nu_e\to\bar\nu_\mu$
oscillations.
The ultimate setup to explore this effect is certainly a neutrino
factory with one or two baselines of several thousand km, see
ref.~\cite{Huber:2006wb} for a recent study. Other options are intense
super beam experiments with a detector at a baseline $L \gtrsim
1000$~km~\cite{wbb,t2kk}, or very long-baseline high gamma beta beam
experiments~\cite{BBhigh}.

In this work we will consider a different strategy, and discuss the
possibility to determine the mass hierarchy with experiments in the
regime of {\it small} matter effect. We will consider experiments with
baselines in the range 100~km~$\lesssim L \lesssim 800$~km and in the
energy range of 300~MeV~$\lesssim E \lesssim$~few~GeV. In this regime
the size of the matter effect is typically a few percent.
We point out that if information from all four CP and T-conjugate
oscillation channels $\nu_\mu\to\nu_e$, $\bar\nu_\mu\to\bar\nu_e$,
$\nu_e\to\nu_\mu$, and $\bar\nu_e\to\bar\nu_\mu$, is available such
small effects can be explored efficiently in order to determine the
neutrino mass hierarchy. The basic observation is that the so-called
sign-degenerate solution~\cite{Minakata:2001qm,Barger:2001yr}, which
prevents the determination of the mass hierarchy, moves in opposite
directions in the plane of $\theta_{13}$ and \delCP\ for T-conjugate
channels. This effect has been observed in ref.~\cite{Campagne:2006yx}
in a simulation of CERN based beta beam and super beam experiments.
Here we discuss the underlying physics and point out the general
principle without specializing to a specific experimental
configuration. This method could be applied for example for the
combination of generic beta beam ($\stackrel{\scriptscriptstyle
(-)}{\nu}_e \to \stackrel{\scriptscriptstyle(-)}{\nu}_\mu$) and super
beam ($\stackrel{\scriptscriptstyle (-)}{\nu}_\mu \to
\stackrel{\scriptscriptstyle(-)}{\nu}_e$) experiments.  Furthermore, a
low energy neutrino factory~\cite{Geer:2007kn} with a detector (or
detectors) capable of muon and electron charge identification would
offer a place to apply this method. If charge identification for
electrons is not possible, the neutrino factory (providing the
$\stackrel{\scriptscriptstyle (-)}{\nu}_e \to
\stackrel{\scriptscriptstyle(-)}{\nu}_\mu$ information) could be
combined with a $\stackrel{\scriptscriptstyle (-)}{\nu}_\mu \to
\stackrel{\scriptscriptstyle(-)}{\nu}_e$ super beam
experiment~\cite{Burguet-Castell:2002qx}.

The outline of the remainder of the paper is as follows. In
sec.~\ref{sec:analytical} we discuss the proposed hierarchy
determination by considerations of the relevant oscillation
probabilities. We discuss the resolution of the sign-degenerate
solution by using analytical expressions for the probabilities and
confirm the results by numerical calculations. In
sec.~\ref{sec:numerical} the method is illustrated with simulations of
realistic experimental configurations. As examples we combine the
$\mu\to e$ super beam experiments SPL, T2HK, and NO$\nu$A with a
$\gamma=100$ beta beam ($e\to\mu$), all operating in the regime of
relatively small matter effect. A summary is presented in
sec.~\ref{sec:summary}.

%%%%%%%%%%%%%%%%%%%%%%%%%%%%%%%%%%%
\section{Resolving the sgn($\Delta m^2_{31}$)-degeneracy with 
T-conjugated oscillation channels}
%%%%%%%%%%%%%%%%%%%%%%%%%%%%%%%%%%%
\label{sec:analytical}

In this section we discuss how the sign-degeneracy can be resolved
with CP and T-conjugated channels by considering the location of the
degenerate solution in the plane of $\theta_{13}$ and \delCP. Similar
considerations can be found for example in
refs.~\cite{Minakata:2001qm,Barger:2001yr,Minakata:2002qi,Donini:2003vz}.
Let us depart from an expression for the appearance oscillation
probability in the $\nu_e-\nu_\mu$ sector, expanded to second order in
the small parameters $s_{13} \equiv \sin\theta_{13}$ and $\alpha
\equiv \Delta m^2_{21} / |\Delta m^2_{31}|$ valid for constant matter
density~\cite{Cervera:2000kp,Freund:2001pn,Akhmedov:2004ny}:
\begin{align}
P_\mathrm{app} &\approx 4 \, s_{13}^2 \, s_{23}^2 \frac{\sin^2
         \Delta (1 - ahA)}{(1 - ahA)^2} +
         \alpha^2 \sin 2\theta_{12} \, c_{23}^2 
         \frac{\sin^2 A\Delta}{A^2} \nonumber\\
         &+ 2 h\, \alpha \, s_{13} \, \sin 2\theta_{12} \, 
         \sin2\theta_{23} \cos(h\Delta - at\delta_{\rm CP}) \, 
	 \frac{\sin\Delta A}{A} \, \frac{\sin \Delta (1 - ahA)}{1-ahA} \,,
\label{eq:P}
\end{align}
with the definitions
\begin{equation}
\Delta \equiv \frac{|\Delta m^2_{31}| L}{4E} \,,\quad
A \equiv \left| \frac{2EV}{\Delta m^2_{31}} \right|\,,
\end{equation}
where $L$ is the baseline, $E$ is the neutrino energy, and $V$ is the
effective matter potential~\cite{Wolfenstein:1977ue}. The signs
$a,t,h$ describe the effects of CP-conjugation, T-conjugation, and the
neutrino mass hierarchy, respectively:
\begin{equation}
a = \left\{ 
  \begin{array}{l}
     +1 \quad \mbox{for $\nu$}  \\
     -1 \quad \mbox{for $\bar\nu$}  
   \end{array}\right.
,\qquad
t = \left\{ 
  \begin{array}{l}
     +1 \quad \mbox{for $e \to \mu$}  \\
     -1 \quad \mbox{for $\mu \to e$}  
   \end{array}\right.
,\qquad
h = \mathrm{sgn}(\Delta m^2_{31}) \,.
\end{equation}
The matter effect enters via the parameter $A$. It is clear from
eq.~(\ref{eq:P}) that in the case of large matter effect $A \gtrsim 1$
the terms $(1 - ahA)$ depend strongly on the type of the mass
hierarchy, and for $ah = 1$ (neutrinos and NH, or anti-neutrinos and
IH) a resonance is encountered for $A = 1$~\cite{Mikheev:1986gs}.

In the following we will focus on a different situation, namely the regime
of small matter effect $A \ll 1$. Numerically one finds for a matter
density of 3~g/cm$^3$
\begin{equation}\label{eq:A}
A \simeq 0.09 \, 
\left(\frac{E}{\rm GeV}\right)
\left(\frac{|\Delta m^2_{31}|}{2.5\times 10^{-3} 
\: \mathrm{eV}^2}\right)^{-1} \,.
\end{equation} 
Furthermore, we concentrate on experiments operating at the first
oscillation maximum, which is characterized by $\Delta \simeq \pi/2$,
or
\begin{equation}\label{eq:max}
E \simeq 0.2 \, {\rm GeV} 
\left(\frac{L}{100 \,\rm km} \right)
\left(\frac{|\Delta m^2_{31}|}{2.5\times 10^{-3} \: \mathrm{eV}^2}\right) \,.
\end{equation}
Eqs.~(\ref{eq:A}) and (\ref{eq:max}) imply that for experiments at
baselines of 130~km, 295~km, 730~km the matter effect is of order 2\%,
5\%, 13\%, respectively.\footnote{A discussion of the issue in which
regions of $L$ and $E$ the matter effect is important can be found
for example in ref.~\cite{Akhmedov:2004ny}.}
Keeping this in mind it makes sense to expand the probability
eq.~(\ref{eq:P}) also in the small quantity $A$. To simplify the
following equations we set $\theta_{23} = \pi/4$ and use the
abbreviation 
\begin{equation}
\alphat \equiv \sin 2\theta_{12} \, \frac{\Delta m^2_{21} L}{4E} \,.
\end{equation}
Then eq.~(\ref{eq:P}) becomes to first order in $A$
\begin{align}
P_\mathrm{app} 
 &\approx 2s^2_{13} \sin^2\Delta  + \frac{1}{2} \, \alphat^2
 + 2h \, \alphat \, s_{13} \, \sin\Delta \cos(h\Delta - at\delCP) \nonumber\\
 &+ 2a \, s_{13}A \, (\sin\Delta - \Delta\cos\Delta)
      \left[2 h\, s_{13}\sin\Delta + \alphat \cos(h\Delta - at\delCP)\right] \,,
      \label{eq:P2}
\end{align}
where the first line is just the vacuum probability and the second
line corresponds to the leading order matter effect correction.

The reason why it is difficult to determine the neutrino mass
hierarchy is a parameter degeneracy with $s_{13}$ and
\delCP~\cite{Minakata:2001qm,Barger:2001yr}, i.e., for given $s_{13}$,
\delCP, and sgn($\Delta m^2_{31}$) in many situations the same
probability can be obtained for the opposite sign of $\Delta m^2_{31}$
and different values $s_{13}'$ and $\delCP'$:
\begin{equation}\label{eq:deg}
P_\mathrm{app}(a,t; h, s_{13}, \delCP) = 
P_\mathrm{app}(a,t; -h, s_{13}', \delCP') \,.
\end{equation}
Assuming a given oscillation channel $t$ and neutrino and anti-neutrino
data ($a = \pm 1$) this is a system of two equations for the variables
$s_{13}'$ and $\delCP'$. If a solution to this system exists the mass
hierarchy cannot be determined.
For example, in the case of vacuum oscillations it follows immediately
from eq.~(\ref{eq:P2}) that the condition (\ref{eq:deg}) can be
fulfilled for $s_{13}' = s_{13}$ and $\delCP' = \pi -
\delCP$~\cite{Minakata:2001qm}:
\begin{equation}\label{eq:degvac}
P_\mathrm{app}^\mathrm{vac}(a,t; h, s_{13}, \delCP) = 
P_\mathrm{app}^\mathrm{vac}(a,t; -h, s_{13}, \pi - \delCP) \,,
\end{equation}
where $P_\mathrm{app}^\mathrm{vac}$ is given by the first line of
eq.~(\ref{eq:P2}), and eq.~(\ref{eq:degvac}) holds independently of
$a,t$ and the neutrino energy $E$. 

To include the leading order matter effect correction we introduce
small deviations from this vacuum solution:
\begin{equation}\label{eq:prime}
s_{13}' = s_{13}(1 + \epsilon_s) \,,\qquad
\delCP' = \pi - \delCP + \epsilon_\delta \,,
\end{equation}
with $\epsilon_s,\epsilon_\delta \ll 1$. Using eqs.~(\ref{eq:P2}) and
(\ref{eq:prime}) in eq.~(\ref{eq:deg}) and expanding to first order in
$\epsilon_s,\epsilon_\delta$, and $A$ yields the condition
\begin{equation}\label{eq:deglin}
\left[\epsilon_s - 2ahA(1-\Delta\cot\Delta)\right]
\left[2h\,s_{13} \sin\Delta + \alphat \cos(h\Delta - at\delCP)\right]
=
at\epsilon_\delta\alphat\sin(h\Delta - at\delCP) \,.
\end{equation}
For a given ``true'' hierarchy $h$, a fixed oscillation channel $t$,
and $a = \pm 1$ this is a linear system of two equations for
$\epsilon_s$ and $\epsilon_\delta$, which in general has a unique
solution. Hence, for neutrino plus anti-neutrino data in one
oscillation channel the leading order matter effect cannot break the
degeneracy. This is the reason why experiments at relatively small
baselines ($L \lesssim 700$~km) have very poor sensitivity to the mass
hierarchy.\footnote{Eq.~(\ref{eq:deglin}) can be fulfilled exactly
only for one energy. Hence, in principle spectral information can be
used to resolve the degeneracy. Note, however, that this is an effect
at third order in the small quantities $\epsilon_s, \epsilon_\delta,
A, s_{13}, \alphat$. Hence, it is difficult to obtain enough
statistics to explore spectral information.} In order to resolve the
degeneracy one has to enter the regime of {\it large} matter effect,
where the non-linear character of eq.~(\ref{eq:P}) becomes relevant
and prevents a solution of the two conditions (\ref{eq:deg}) for fixed
$t$ and $a\pm 1$.

However, the immediate conclusion from eq.~(\ref{eq:deglin}) is that
if all CP and T-conjugate channels are available one obtains four
independent relations (corresponding to $a = \pm 1$ and $t = \pm 1$)
for the two variables $\epsilon_s$ and $\epsilon_\delta$. Obviously,
in general such a system has no solution and hence the degeneracy is
broken. 
To illustrate this explicitly let us consider for simplicity the case
$\Delta = \pi / 2$, i.e., experiments exactly at the first oscillation
maximum. Then eq.~(\ref{eq:deglin}) simplifies to
\begin{equation}\label{eq:deglinmax}
(\epsilon_s - 2ahA)(2s_{13} + at \,\alphat \sin\delCP)
=
at \, \epsilon_\delta \, \alphat\cos\delCP \,.
\end{equation}
For $a = \pm 1$ and given $t$ this system of equations has the solution
\begin{align}
\epsilon_s &= ht \, A \, \frac{\alphat}{s_{13}} \, \sin\delCP \,,
\nonumber\\
\epsilon_\delta &= ht \, \frac{A}{\cos\delCP} \, 
\left(\frac{\alphat}{s_{13}} \, \sin^2\delCP -
4 \frac{s_{13}}{\alphat}\right) \,. 
\label{eq:solution}
\end{align}
The crucial observation from these expressions is that the signs of
both, $\epsilon_s$ and $\epsilon_\delta$, depend on the oscillation
channel $t$. Hence, with increasing matter effect $A$ the location of
the solution with the wrong hierarchy in the $\theta_{13}-\delCP$
plane moves in opposite directions for $\mu\to e$ and $e\to\mu$
transitions.

\begin{figure}[!t]
  \centering \includegraphics[width=0.8\textwidth]{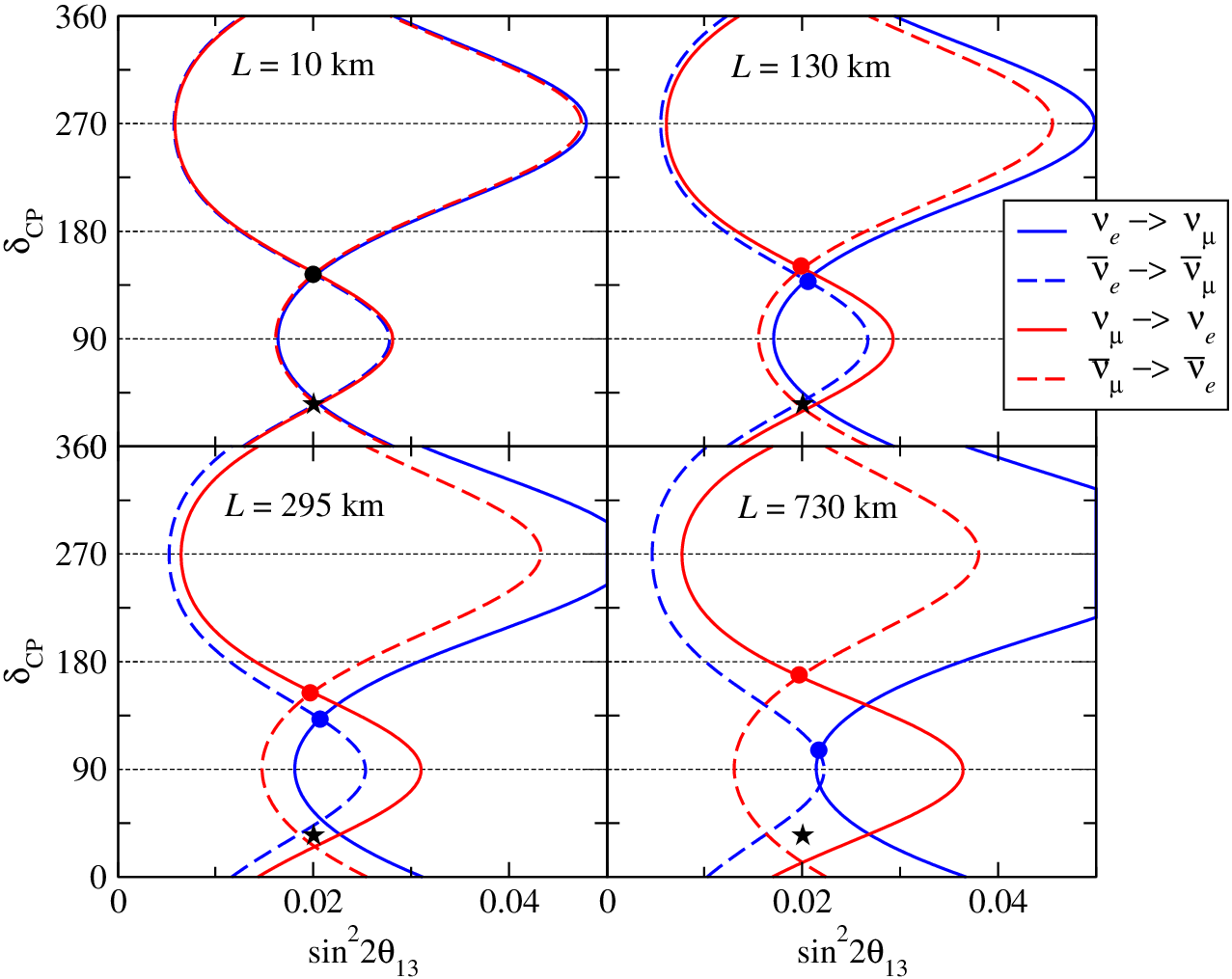}
  \mycaption{Location of the sign-degenerate solutions for different
  baselines. The plots show a graphical representation of the
  equations $P_\mathrm{app}(a,t; {\rm NH}, s_{13}, \delCP) =
  P_\mathrm{app}(a,t; {\rm IH}, s_{13}', \delCP')$ for $a = \pm 1$
  (neutrinos/anti-neutrinos) and $t = \pm 1$ ($(e\to\mu)/(\mu\to e)$). The
  star indicates the assumed values for \stheta\ and \delCP\ for the
  NH, whereas the axes correspond to the primed parameters for the
  wrong hierarchy. The dots show the location of the degenerate
  solutions, where the curves for neutrinos and anti-neutrinos for fixed
  $t$ cross. The four panels correspond to different baselines, and
  the neutrino energy in each panel is determined by assuming the
  first oscillation maximum and $|\Delta m^2_{31}| = 2.5\times
  10^{-3}$~eV$^2$, see eq.~(\ref{eq:max}). \label{fig:P}}
\end{figure}

We have verified this behaviour by numerical calculations of the
oscillation probability for constant matter (without any expansion in
small quantities). Based on such calculations we graphically solve the
system of equations (\ref{eq:deg}) in fig.~\ref{fig:P}. The appearance
probability is calculated for NH and the parameters $\stheta = 0.02$
and $\delCP = 36^\circ$ (marked as a star). Then the curves in
fig.~\ref{fig:P} correspond to the set of values $\stheta'$ and
$\delCP'$ leading to the same probability but for IH. In each panel
there are four curves, corresponding to the four combinations of
neutrinos/anti-neutrinos and $(e\to\mu)/(\mu\to e)$. The dots show the
location of the degenerate solutions, where the curves for neutrinos
and anti-neutrinos for a given channel cross. The second crossing
close to the original parameter values correspond to the so-called
intrinsic degenerate
solution~\cite{Burguet-Castell:2001ez,Barger:2001yr} (with the wrong
hierarchy). For experiments at the first oscillation maximum this
degeneracy is resolved quite efficiently by spectral information, and
hence, in many cases it does not appear as a viable solution (see,
e.g., ref.~\cite{Huber:2002mx} for an explicit discussion in the case
of the T2HK experiment). Therefore we will neglect it in the present
discussion and focus on the solutions marked with dots in
fig.~\ref{fig:P}.

In the upper-left panel of fig.~\ref{fig:P} we consider a
(hypothetical) experiment at a very short baseline $L = 10$~km and an
energy of 0.02~GeV, where the matter effect is negligible, see
eq.~(\ref{eq:A}). One finds that all four curves meet in the point
corresponding to eq.~(\ref{eq:degvac}) with $\delCP' = \pi - \delCP$,
which makes the determination of the hierarchy impossible even if all
CP and T-conjugated channels are available. By increasing the baseline
(and simultaneously choosing the energy to stay in the first
oscillation maximum) one observes from the plots that the degenerate
solutions for the $e\to\mu$ and $\mu\to e$ channels separate and move
in opposite directions, in agreement with eq.~(\ref{eq:solution}).

%%%%%%%%%%%%%%%%%%%%%%%%%%%%%%%%%%%%%%%%%%%%%
\section{Numerical simulations}
%%%%%%%%%%%%%%%%%%%%%%%%%%%%%%%%%%%%%%%%%%%%%
\label{sec:numerical}

\begin{table}
\centering
\begin{tabular}{|lccccllc|}
\hline
Exp. & Ref. & $L$ [km] & $\langle E\rangle$ [GeV] & Detector & Time
[yr] & Beam & $\sigma_\mathrm{sys}$\\
\hline
BB & \cite{Campagne:2006yx} & 130 & 0.4 & 500 kt WC & $4\nu + 4\bar\nu$ & $2.2\, (5.8) \times 10^{18}$ & 2\%\\
SPL& \cite{Campagne:2006yx} & 130 & 0.3 & 500 kt WC & $2\nu + 8\bar\nu$ & 4 MW & 2\%\\
T2HK & \cite{T2K} & 295 & 0.8 &  500 kt WC & $4\nu + 4\bar\nu$ & 4 MW & 5\%\\
NO$\nu$A & \cite{Ayres:2004js} & 812 & 2.0 & 25 kt TASD & $3\nu + 3\bar\nu$ & 1.12 MW & 5\%\\
\hline
\end{tabular}
  \mycaption{Main parameters of the simulated
  setups~\cite{globes}. For BB the ``beam'' column corresponds to the
  number of useful \Ne\ (\He) decays per year, whereas for the super
  beams the beam power is given. The systematical error
  $\sigma_\mathrm{sys}$ corresponds to the uncertainty on the signal
  and background rates, uncorrelated between signal, background,
  neutrinos, and anti-neutrinos.
\label{tab:setups}}
\end{table}

In this section we show by numerical simulations of realistic
experimental configurations how one can benefit from the synergy of
T-conjugated oscillation channels. We consider the three super beam
experiments SPL, T2HK, and NO$\nu$A, providing the
$\stackrel{\scriptscriptstyle (-)}{\nu}_\mu \to
\stackrel{\scriptscriptstyle(-)}{\nu}_e$ information, as well as a
$\gamma = 100$ beta beam (BB) experiment operating in the
$\stackrel{\scriptscriptstyle (-)}{\nu}_e \to
\stackrel{\scriptscriptstyle(-)}{\nu}_\mu$ channels. The simulation is
performed with the GLoBES software~\cite{globes}, and all setups
correspond to the pre-defined configurations provided by GLoBES~3.0.
The most relevant parameters for each experiment are given in
tab.~\ref{tab:setups}.
BB and SPL are CERN based experiments, using the 500~kt water
\v{C}erenkov (WC) detector MEMPHYS at Frejus, at a distance of 130~km,
details can be found in ref.~\cite{Campagne:2006yx}. With such a short
baseline and low energies the matter effect is very small. T2HK is the
second phase of the T2K experiment~\cite{T2K} in Japan, based on a
4~MW upgrade of the beam and the 500~kt HyperKamiokande
detector. Further details of the simulation can be found also in
ref.~\cite{Huber:2002mx}. In this experiment the matter effect is
somewhat larger than for the CERN--MEMPHYS configuration, but still
too small to explore the neutrino mass hierarchy. Finally, we consider
the Fermilab based NO$\nu$A experiment~\cite{Ayres:2004js} with a
25~kt totally active scintillator detector (TASD) at a baseline of
812~km, where the matter effect starts to be important. Note however,
that here we consider an initial stage setup for NO$\nu$A, with
significantly less statistics than the other configurations.

To simulate the data we take the following values for the oscillation
parameters: $\Delta m^2_{21} = 7.9\times 10^{-5}$~eV$^2$,
$\sin^2\theta_{12} = 0.3$, $\Delta m^2_{31} = +2.4\times
10^{-3}$~eV$^2$, $\sin^2\theta_{23} = 0.5$, and we assume an external
uncertainty of 4\% on the solar parameters, and 5\% on the matter
density uncertainty along the baseline of each experiment. Note that
each experiment includes neutrino and anti-neutrino data, and 
appearance and disappearance channels are used in the analysis.

\begin{figure}[!t]
  \centering \includegraphics[width=0.6\textwidth]{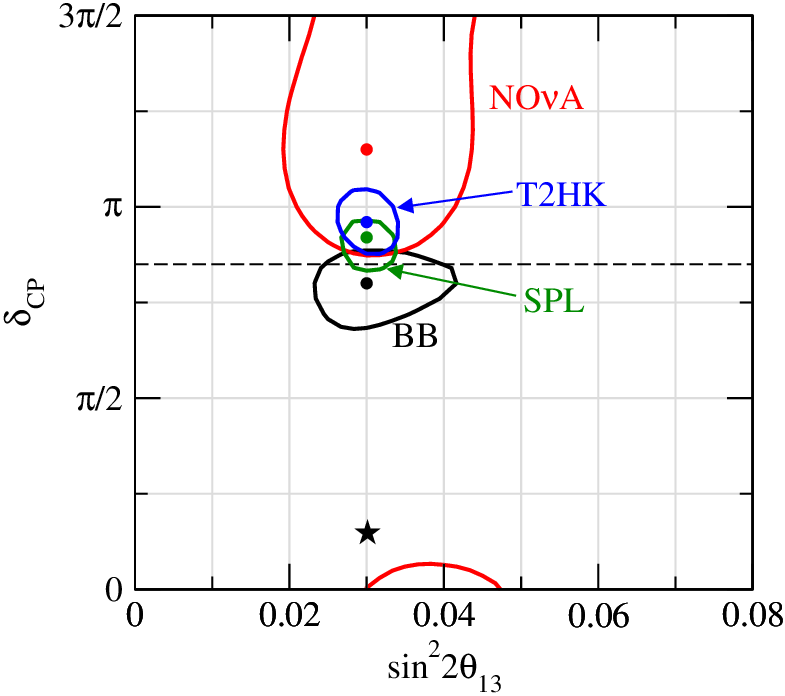}
  \mycaption{Allowed regions at 90\%~CL in the $\stheta-\delCP$ plane
  with the wrong hierarchy for BB, SPL, T2HK, and NO$\nu$A. Data is
  simulated for NH and $\stheta^\mathrm{true} = 0.03$,
  $\delCP^\mathrm{true} = 0.15\pi$, marked with a star. The dots show
  the best fit point with IH hierarchy. The horizontal dashed line
  indicates the value $\delCP = \pi - \delCP^\mathrm{true}$,
  corresponding to the location of the degenerate solution in vacuum,
  see eq.~(\ref{eq:degvac}). For the BB external accuracies on
  $|\Delta m^2_{31}|$ and $\theta_{23}$ of 5\% and 10\% have been
  assumed, respectively.
  \label{fig:d-th}}
\end{figure}

In fig.~\ref{fig:d-th} we show an example of the sign-degenerate
solutions in the $\stheta-\delCP$ plane for the four experiments of
tab.~\ref{tab:setups}. This plot confirms the behaviour discussed in
the previous section on probability level by performing an actual fit
to simulated data: One can see that the best fit point with the wrong
hierarchy moves in opposite directions for BB ($e\to\mu$) and the super
beam experiments ($\mu\to e$), relative to the vacuum solution at
$\delCP = \pi - \delCP^\mathrm{true}$, which is indicated by the
dashed line. Furthermore, the dislocation of the degeneracy for the
super beam experiments from the vacuum value of \delCP\ is
proportional to the baseline, i.e., increasing in the order SPL, T2HK,
NO$\nu$A.  This behaviour shows that the combination of $\mu\to e$ and
$e\to\mu$ experiments offers a promising synergy in resolving the
sign-degeneracy. 
Note that that in all cases only one solution appears with the wrong
hierarchy, which justifies to neglect the intrinsic degeneracy in the
discussion of the previous section.

Let us stress that in each case a detailed quantitative study is
necessary in order to fully assess the potential of this method. The
final ability to disfavour the degenerate solution depends on many
details which affect the size of the allowed regions for the
individual experiments. For example, the relatively large allowed
region for NO$\nu$A seen in fig.~\ref{fig:d-th}, which is a
consequence of the much smaller statistics compared to the other
setups, will certainly limit the sensitivity, tough the best fit point
of the degeneracy for NO$\nu$A clearly follows the trend discussed
above. For the SPL/BB combination the effect discussed here has
apparently not be found in
refs.~\cite{Donini:2003vz,Donini:2004hu}. It is clear from
fig.~\ref{fig:d-th} that for SPL+BB a slight enlargement of the
allowed regions would be enough to corrupt the ability to resolve the
sign-degeneracy. The size of the allowed regions depends rather
sensitively on the details of the analysis (see e.g., fig.~6 of
ref.~\cite{Campagne:2006yx} for the case of SPL). A possible reason
for the differences with refs.~\cite{Donini:2003vz,Donini:2004hu}
might be for example the assumptions on systematics (see
tab.~\ref{tab:setups}), or the inclusion of spectral information for
appearance as well as disappearance channels, which has quite a
significant impact on the size of the allowed
regions~\cite{Campagne:2006yx}.

In fig.~\ref{fig:d-th} we have assumed a ``true'' NH. If the true
hierarchy is inverted the degenerate solutions for the $\mu\to e$ and
$e\to\mu$ channels move into the opposite directions as in the case of
a true NH. This follows from eq.~(\ref{eq:solution}), where $h$
corresponds to the true sign of $\Delta m^2_{31}$. Of course, the
complementarity between the T-conjugated channels remains
independently of the true hierarchy.

Let us comment on the widely discussed strategy of using information
from two $\mu\to e$ experiments at similar $L/E$ but different
baselines, see, e.g., ref.~\cite{Barger:2002xk}. Such a situation is
realized by the combination of T2K and NO$\nu$A~\cite{t2k+nova} or by
placing a second detector in the NO$\nu$A beam-line at a suitable
off-axis angle, as proposed in ref.~\cite{MenaRequejo:2005hn}. It is
clear from the preceding discussion and from fig.~\ref{fig:d-th} that
the synergy from such a combination is less effective than using
$e\to\mu$ information. The reason is that the degeneracies for two
$\mu\to e$ experiments move in the {\it same} direction in the
$\stheta-\delCP$ space, only the size of the dislocation is different
due to the different baselines.\footnote{Note that the present
discussion does not necessarily apply to super beam experiments with
neutrino energies $E \sim$~GeV and a detector at large baselines $L
\gtrsim 1000$~km, far beyond the first oscillation
maximum~\cite{wbb,t2kk}. Apart from the fact that in this case the
matter effect cannot be considered to be small, also the rich
information from the energy spectrum at higher oscillation maxima can
lead to a rather different behaviour of the sign-degenerate solution
than discussed here.}

\begin{figure}[!t]
  \centering \includegraphics[width=0.9\textwidth]{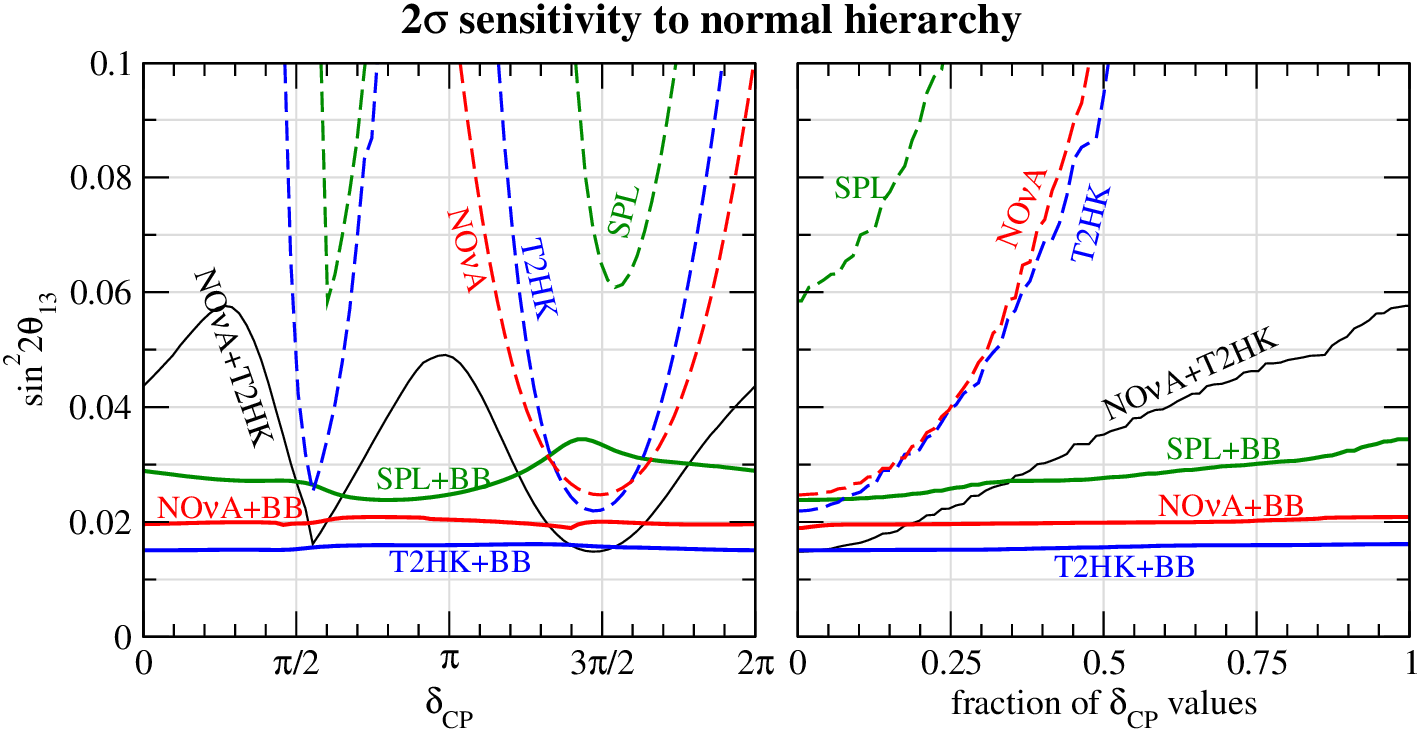}
  \mycaption{Minimal value of $\stheta$ for which the IH can be
  excluded at $2\sigma$ ($\Delta\chi^2 = 4$) if the true hierarchy is
  normal, as a function of \delCP\ (left) and the fraction of all
  possible values of \delCP\ (right). The dashed lines correspond to
  the super beam experiments SPL, T2HK, NO$\nu$A, whereas for the
  solid lines each of these super beams is combined with the beta
  beam. For comparison also the combination NO$\nu$A+T2HK is shown.
\label{fig:sens}}
\end{figure}

In fig.~\ref{fig:sens} we show how the combination of the super beam
experiments SPL, T2HK, and NO$\nu$A with the $\gamma=100$ beta beam
significantly enhances the sensitivity to the mass hierarchy due to
the $(\mu\to e)/(e\to\mu)$ synergy. The dashed curves show the
sensitivities of the super beams alone. One observes that these
experiments can assess the mass hierarchy only in a certain range of
\delCP\ values, and there is no sensitivity even for $\stheta = 0.1$
for 75\% (50\%) of all values of \delCP\ for SPL (T2HK,
NO$\nu$A). Note that NO$\nu$A and T2HK have a rather similar
sensitivity, where NO$\nu$A has the advantage of the longer baseline
of 812~km, whereas the short baseline of T2HK of 295~km is compensated
by the large statistics implied by the 4~MW beam and
500~kt detector. The main limitation of SPL is of course the short
baseline of 130~km.

When these experiments are combined with the BB the sensitivity is
significantly improved, see solid lines in fig.~\ref{fig:sens}. The
dependence on \delCP\ practically disappears and a stable sensitivity
is obtained for $\stheta \gtrsim 0.02-0.03$.\footnote{In
fig.~\ref{fig:sens} we have assumed that the true hierarchy is normal,
but we have checked that for a true IH the results are very similar.}
The effect is most remarkable for SPL+BB~\cite{Campagne:2006yx}, since
none of these experiments on its own has any notable
sensitivity. Indeed, in the parameter range shown in
fig.~\ref{fig:sens} there is no sensitivity for BB alone. The
sensitivity of the SPL+BB combination fully emerges from the
complementarity of T-conjugated channels in the small matter effect
regime. Also for T2HK and NO$\nu$A the effect is clearly visible, and
again the larger baseline of NO$\nu$A is compensated by statistics in
T2HK.
For comparison we show also the combination NO$\nu$A+T2HK (without BB)
in fig.~\ref{fig:sens} (see ref.~\cite{t2k+nova} for detailed
discussions of the case NO$\nu$A + T2K phase~I). Also in this case the
sensitivity improves, however the complementarity is much less
pronounced, a dependence on \delCP\ remains, and the effect is more
similar to the addition of statistics than a true synergy (see also
the corresponding discussion related to fig.~\ref{fig:d-th}).

As a side remark let us mention the possibility pointed out in the
second paper of ref.~\cite{non-matter}, that very accurate
measurements of the neutrino mass-squared difference in $\nu_e$ and
$\nu_\mu$ disappearance experiments allow in principle to distinguish
between NH and IH (even in vacuum). In the beta beam/super beam
combination both disappearance probabilities $P_{ee}$ and $P_{\mu\mu}$
are observed. However, we have checked that for the experiments
considered here numerically this effect is completely negligible, and
practically the full sensitivity shown in fig.~\ref{fig:sens} emerges
from the matter effect in the appearance channels.

%%%%%%%%%%%%%%%%%%%%%%%%%%%%%%%%%%%%%
\section{Summary and discussion}
%%%%%%%%%%%%%%%%%%%%%%%%%%%%%%%%%%%%%
\label{sec:summary}

In this letter we have considered neutrino oscillation experiments
operating at the first oscillation maximum in the regime of small
matter effect, i.e., at relatively short baselines of several hundred
km. In such a case there is very poor sensitivity to the neutrino mass
hierarchy, if only data from neutrinos and anti-neutrinos are available
in a fixed oscillation channel. The reason is that the leading order
correction in the small matter effect parameter $A$ cannot break the
sgn($\Delta m^2_{31}$)-degeneracy. The usual strategy to resolve the
degeneracy is to enter the regime of strong matter effect (by going to
longer baselines and higher neutrino energies), where the non-linear
dependence of $A$ becomes important.  
Here we have proposed an alternative method, namely the combination of
all four CP and T-conjugated oscillation channels. We have shown that
the location of the sign-degenerate solution in the plane of
$\theta_{13}$ and \delCP\ moves in opposite directions for the 
$\stackrel{\scriptscriptstyle (-)}{\nu}_\mu \to
\stackrel{\scriptscriptstyle(-)}{\nu}_e$ and
$\stackrel{\scriptscriptstyle (-)}{\nu}_e \to
\stackrel{\scriptscriptstyle(-)}{\nu}_\mu$ channels when the matter
effect increases. This synergy allows to resolve the degeneracy even
for matter effects as small as a few percent.
We have discussed the method at the level of oscillation
probabilities, and illustrated the effect also by simulations of
representative experimental configurations.

A typical situation where our method applies is the combination of 
beta beam ($e\to\mu$) and super beam ($\mu\to e$) experiments. We have
demonstrated that a significant synergy exists for the determination
of the mass hierarchy by simulations of the SPL, T2HK, and NO$\nu$A
super beams combined with a $\gamma = 100$ beta beam. Most remarkable,
even for the SPL/beta beam combination, where both experiments have a
baseline of only 130~km, there is sensitivity to the mass hierarchy
for $\stheta \gtrsim 0.03$ due to the synergy of the T-conjugated
channels. 
Another possibility to take advantage of this effect could be a low
energy neutrino factory~\cite{Geer:2007kn} operating in an $L/E$
regime where the matter effect is not yet fully developed. The liquid
argon technology, which has been considered for the detector in that
reference, has excellent sensitivity for muon as well as electron
detection. If in both cases the charge can be identified all four
oscillation channels $\stackrel{\scriptscriptstyle (-)}{\nu}_\mu \to
\stackrel{\scriptscriptstyle(-)}{\nu}_e$ and
$\stackrel{\scriptscriptstyle (-)}{\nu}_e \to
\stackrel{\scriptscriptstyle(-)}{\nu}_\mu$ were available at the
neutrino factory. If charge identification is not possible for
electrons one could combine the neutrino factory with a super beam
experiment providing the $\mu\to e$ information.

The purpose of this note is to point out the existence of a synergy of
T-conjugated channels, which can significantly increase the
sensitivity to the neutrino mass hierarchy. Whether this method for
the mass hierarchy determination is indeed useful for a given
experimental configuration, or is competitive with alternative
approaches needs to be confirmed by detailed simulations and
comparison studies, which is beyond the scope of this work.

\renewcommand{\baselinestretch}{1}

%\newpage

%%%%%%%%%%%%%%%%%%%%%%%%%%%%%%%%%%%%%%%%%%%%%%%%%%%%

\begin{thebibliography}{99}

\bibitem{Wolfenstein:1977ue}
  L.~Wolfenstein,
  %{\it Neutrino oscillations in matter,}
  Phys.\ Rev.\  D {\bf 17} (1978) 2369;
  %%CITATION = PHRVA,D17,2369;%%
%
%\bibitem{Wolfenstein:1979ni}
%  L.~Wolfenstein,
  %{\it Neutrino Oscillations And Stellar Collapse,}
  Phys.\ Rev.\  D {\bf 20} (1979) 2634.
  %%CITATION = PHRVA,D20,2634;%%

\bibitem{Barger:1980tf}
  V.~D.~Barger, K.~Whisnant, S.~Pakvasa and R.~J.~N.~Phillips,
  %{\it Matter effects on three-neutrino oscillations,}
  Phys.\ Rev.\  D {\bf 22} (1980) 2718.
  %%CITATION = PHRVA,D22,2718;%%


\bibitem{Mikheev:1986gs}
  S.~P.~Mikheev and A.~Y.~Smirnov,
  %{\it Resonance enhancement of oscillations in matter and solar neutrino
  %spectroscopy,}
  Sov.\ J.\ Nucl.\ Phys.\  {\bf 42} (1985) 913
  [Yad.\ Fiz.\  {\bf 42} (1985) 1441];
  %%CITATION = YAFIA,42,1441;%%
%  
%\bibitem{Mikheev:1986wj}
%  S.~P.~Mikheev and A.~Y.~Smirnov,
  %{\it Resonant amplification of neutrino oscillations in matter and solar
  %neutrino spectroscopy,}
  Nuovo Cim.\  C {\bf 9} (1986) 17.
  %%CITATION = NUCIA,9C,17;%%  

\bibitem{Freund:1999gy}
  M.~Freund, M.~Lindner, S.T.~Petcov and A.~Romanino,
  %{\it Testing matter effects in very long baseline neutrino oscillation
  %experiments,}
  Nucl.\ Phys.\ B {\bf 578}, 27 (2000)
  [hep-ph/9912457];
%
%\bibitem{Barger:2000cp}
  V.D.~Barger, S.~Geer, R.~Raja and K.~Whisnant,
  %{\it Determination of the pattern of neutrino masses at a neutrino factory,}
  Phys.\ Lett.\ B {\bf 485}, 379 (2000)
  [hep-ph/0004208].
  %%CITATION = HEP-PH 0004208;%%

\bibitem{atm}
%
%\bibitem{Akhmedov:1998xq}
  E.~K.~Akhmedov, A.~Dighe, P.~Lipari and A.~Y.~Smirnov,
  %{\it Atmospheric neutrinos at Super-Kamiokande and parametric resonance in
  %neutrino oscillations,}
  Nucl.\ Phys.\ B {\bf 542}, 3 (1999)
  [hep-ph/9808270];
  %%CITATION = HEP-PH 9808270;%%
%
%\bibitem{Bernabeu:2003yp}
  J.~Bernabeu, S.~Palomares-Ruiz and S.~T.~Petcov,
  %{\it Atmospheric neutrino oscillations, $\theta_{13}$ and neutrino mass
  %hierarchy,}
  Nucl.\ Phys.\ B {\bf 669}, 255 (2003)
  [hep-ph/0305152];
  %%CITATION = HEP-PH 0305152;%%
%
%\bibitem{Indumathi:2004kd}
  D.~Indumathi and M.~V.~N.~Murthy,
  %{\it A question of hierarchy: Matter effects with atmospheric neutrinos and
  %anti-neutrinos,}
  Phys.\ Rev.\  D {\bf 71} (2005) 013001
  [hep-ph/0407336];
  %%CITATION = PHRVA,D71,013001;%%
%
%\bibitem{Huber:2005ep}
  P.~Huber, M.~Maltoni, T.~Schwetz,
  %{\it Resolving parameter degeneracies in long-baseline experiments by
  %atmospheric neutrino data,}
  Phys.\ Rev.\ D {\bf 71} (2005) 053006
  [hep-ph/0501037];
  %%CITATION = HEP-PH 0501037;%%
%
%\bibitem{Petcov:2005rv}
  S.~T.~Petcov and T.~Schwetz,
  %{\it Determining the neutrino mass hierarchy with atmospheric neutrinos,}
  Nucl.\ Phys.\ B {\bf 740}, 1 (2006)
  [hep-ph/0511277].
  %%CITATION = HEP-PH 0511277;%%

\bibitem{supernova}
%\bibitem{Lunardini:2003eh}
  C.~Lunardini and A.~Y.~Smirnov,
  %{\it Probing the neutrino mass hierarchy and the 13-mixing with supernovae,}
  JCAP {\bf 0306} (2003) 009
  [hep-ph/0302033];
  %%CITATION = JCAPA,0306,009;%%
%
%\bibitem{Dighe:2003be}
  A.~S.~Dighe, M.~T.~Keil and G.~G.~Raffelt,
  %{\it Detecting the neutrino mass hierarchy with a supernova at IceCube,}
  JCAP {\bf 0306} (2003) 005
  [hep-ph/0303210];
  %%CITATION = JCAPA,0306,005;%%
%
%\bibitem{Kachelriess:2004vs}
  M.~Kachelriess and R.~Tomas,
  %{\it Identifying the neutrino mass hierarchy with supernova neutrinos,}
  hep-ph/0412100;
  %%CITATION = HEP-PH/0412100;%%
%
%\bibitem{Barger:2005it}
  V.~Barger, P.~Huber and D.~Marfatia,
  %{\it Supernova neutrinos can tell us the neutrino mass hierarchy  independently
  %of flux models,}
  Phys.\ Lett.\  B {\bf 617} (2005) 167
  [hep-ph/0501184].
  %%CITATION = PHLTA,B617,167;%%

\bibitem{non-matter}
%\bibitem{Petcov:2001sy}
  S.T.~Petcov and M.~Piai,
  %{\it The LMA MSW solution of the solar neutrino problem, inverted neutrino  mass
  %hierarchy and reactor neutrino experiments,}
  Phys.\ Lett.\ B {\bf 533} (2002) 94
  [hep-ph/0112074];
  %%CITATION = HEP-PH 0112074;%%
%
%\bibitem{Nunokawa:2005nx}
  H.~Nunokawa, S.~J.~Parke and R.~Zukanovich Funchal,
  %{\it Another possible way to determine the neutrino mass hierarchy,}
  Phys.\ Rev.\  D {\bf 72} (2005) 013009
  [hep-ph/0503283];
  %%CITATION = PHRVA,D72,013009;%%
%
%\bibitem{deGouvea:2005hk}
  A.~de Gouvea, J.~Jenkins and B.~Kayser,
  %{\it Neutrino mass hierarchy, vacuum oscillations, and vanishing $U_{e3}$,}
  Phys.\ Rev.\  D {\bf 71} (2005) 113009
  [hep-ph/0503079];
  %%CITATION = PHRVA,D71,113009;%%
%
%\bibitem{deGouvea:2005mi}
  A.~de Gouvea and W.~Winter,
  %{\it What would it take to determine the neutrino mass hierarchy if $\theta_{13}$
  %were too small?,}
  Phys.\ Rev.\  D {\bf 73} (2006) 033003
  [hep-ph/0509359].
  %%CITATION = PHRVA,D73,033003;%%

\bibitem{DBD}
%\bibitem{Pascoli:2005zb}
  S.~Pascoli, S.~T.~Petcov and T.~Schwetz,
  %{\it The absolute neutrino mass scale, neutrino mass spectrum, Majorana
  %CP-violation and neutrinoless double-beta decay,}
  Nucl.\ Phys.\  B {\bf 734}, 24 (2006)
  [hep-ph/0505226];
  %%CITATION = NUPHA,B734,24;%%
%
%\bibitem{Choubey:2005rq}
  S.~Choubey and W.~Rodejohann,
  %{\it Neutrinoless double beta decay and future neutrino oscillation  precision
  %experiments,}
  Phys.\ Rev.\  D {\bf 72} (2005) 033016
  [hep-ph/0506102];
  %%CITATION = PHRVA,D72,033016;%%
%
%\bibitem{deGouvea:2005hj}
  A.~de Gouvea and J.~Jenkins,
  %{\it Non-oscillation probes of the neutrino mass hierarchy and vanishing
  %$|U_{e3}|$,}
  hep-ph/0507021.
  %%CITATION = HEP-PH/0507021;%%

\bibitem{Gandhi:2004bj}
  R.~Gandhi, P.~Ghoshal, S.~Goswami, P.~Mehta and S.~Uma Sankar,
  %``Earth matter effects at very long baselines and the neutrino mass
  %hierarchy,''
  Phys.\ Rev.\  D {\bf 73} (2006) 053001
  [hep-ph/0411252].
  %%CITATION = PHRVA,D73,053001;%%

\bibitem{Huber:2006wb}
  P.~Huber, M.~Lindner, M.~Rolinec and W.~Winter,
  %{\it Optimization of a neutrino factory oscillation experiment,}
  Phys.\ Rev.\  D {\bf 74} (2006) 073003
  [hep-ph/0606119].
  %%CITATION = PHRVA,D74,073003;%%

\bibitem{wbb}
%\bibitem{Diwan:2003bp}
  M.~V.~Diwan et al.,
  %{\it Very long baseline neutrino oscillation experiments for precise
  %measurements of mixing parameters and CP violating effects,}
  Phys.\ Rev.\ D {\bf 68} (2003) 012002
  [hep-ph/0303081];
  %%CITATION = HEP-PH 0303081;%%
%
%\bibitem{Barger:2006vy}
  V.~Barger et al., 
  %M.~Dierckxsens, M.~Diwan, P.~Huber, C.~Lewis, D.~Marfatia and B.~Viren,
  %{\it Precision physics with a wide band super neutrino beam,}
  Phys.\ Rev.\  D {\bf 74} (2006) 073004
  [hep-ph/0607177].
  %%CITATION = PHRVA,D74,073004;%%

\bibitem{t2kk}
%\bibitem{Ishitsuka:2005qi}
  M.~Ishitsuka, T.~Kajita, H.~Minakata and H.~Nunokawa,
  %{\it Resolving neutrino mass hierarchy and CP degeneracy by two identical
  %detectors with different baselines,}
  Phys.\ Rev.\ D {\bf 72} (2005) 033003
  [hep-ph/0504026];
  %%CITATION = HEP-PH 0504026;%%
%
%\bibitem{Kajita:2006bt}
  T.~Kajita, H.~Minakata, S.~Nakayama and H.~Nunokawa,
  %{\it Resolving eight-fold neutrino parameter degeneracy by two identical
  %detectors with different baselines,}
  Phys.\ Rev.\  D {\bf 75}, 013006 (2007)
  [hep-ph/0609286];
  %%CITATION = PHRVA,D75,013006;%%
%
%\bibitem{Hagiwara:2005pe}
  K.~Hagiwara, N.~Okamura and K.~Senda,
  %{\it Solving the neutrino parameter degeneracy by measuring the T2K off-axis
  %beam in Korea,}
  Phys.\ Lett.\  B {\bf 637} (2006) 266
  [Erratum-ibid.\  B {\bf 641} (2006) 486]
  [hep-ph/0504061].
  %%CITATION = PHLTA,B637,266;%%

\bibitem{BBhigh}
%\bibitem{Burguet-Castell:2003vv}
  J.~Burguet-Castell, D.~Casper, J.~J.~Gomez-Cadenas, P.~Hernandez and F.~Sanchez,
  %``Neutrino oscillation physics with a higher gamma beta-beam,''
  Nucl.\ Phys.\  B {\bf 695} (2004) 217
  [hep-ph/0312068];
  %%CITATION = NUPHA,B695,217;%%
%
%\bibitem{Huber:2005jk}
  P.~Huber, M.~Lindner, M.~Rolinec and W.~Winter,
  %``Physics and optimization of beta-beams: From low to very high gamma,''
  Phys.\ Rev.\  D {\bf 73} (2006) 053002
  [hep-ph/0506237];
  %%CITATION = PHRVA,D73,053002;%%
%
%\bibitem{Agarwalla:2006vf} 
  S.~K.~Agarwalla, S.~Choubey and A.~Raychaudhuri, 
  %{\it Neutrino mass hierarchy and $\theta_{13}$ with
  %a magic baseline beta-beam experiment,} 
  hep-ph/0610333.  
  %%CITATION = HEP-PH/0610333;%%

\bibitem{Minakata:2001qm}
  H.~Minakata and H.~Nunokawa,
  %{\it Exploring neutrino mixing with low energy superbeams,}
  JHEP {\bf 0110}, 001 (2001)
  [hep-ph/0108085].
  %%CITATION = HEP-PH 0108085;%%

\bibitem{Barger:2001yr}
  V.~Barger, D.~Marfatia and K.~Whisnant,
  %{\it Breaking eight-fold degeneracies in neutrino CP violation, mixing, and
  %mass hierarchy,}
  Phys.\ Rev.\ D {\bf 65}, 073023 (2002)
  [hep-ph/0112119].
  %%CITATION = HEP-PH 0112119;%%

\bibitem{Campagne:2006yx}
  J.~E.~Campagne, M.~Maltoni, M.~Mezzetto and T.~Schwetz,
  %{\it Physics potential of the CERN--MEMPHYS neutrino oscillation project,}
  JHEP, to appear 
  [hep-ph/0603172].
  %%CITATION = HEP-PH/0603172;%%

\bibitem{Geer:2007kn}
  S.~Geer, O.~Mena and S.~Pascoli,
  %{\it A low energy neutrino factory for large $\theta_{13}$,}
  hep-ph/0701258.
  %%CITATION = HEP-PH/0701258;%%

\bibitem{Burguet-Castell:2002qx}
  J.~Burguet-Castell, M.~B.~Gavela, J.~J.~Gomez-Cadenas, P.~Hernandez and O.~Mena,
  %{\it Superbeams plus neutrino factory: The golden path to leptonic CP
  %violation,}
  Nucl.\ Phys.\  B {\bf 646} (2002) 301
  [hep-ph/0207080].
  %%CITATION = NUPHA,B646,301;%%

\bibitem{Minakata:2002qi}
  H.~Minakata, H.~Nunokawa and S.~J.~Parke,
  %{\it Parameter degeneracies in neutrino oscillation measurement of leptonic  CP
  %and T violation,}
  Phys.\ Rev.\  D {\bf 66} (2002) 093012
  [hep-ph/0208163].
  %%CITATION = PHRVA,D66,093012;%%

\bibitem{Donini:2003vz}
  A.~Donini, D.~Meloni and S.~Rigolin,
  %{\it Clone flow analysis for a theory inspired neutrino experiment
  %planning,}
  JHEP {\bf 0406} (2004) 011
  [hep-ph/0312072].
  %%CITATION = JHEPA,0406,011;%%

\bibitem{Cervera:2000kp}
  A.~Cervera et al., 
  %A.~Donini, M.~B.~Gavela, J.~J.~Gomez Cadenas, P.~Hernandez, O.~Mena and S.~Rigolin,
  %{\it Golden measurements at a neutrino factory,}
  Nucl.\ Phys.\  B {\bf 579} (2000) 17
  [Erratum-ibid.\  B {\bf 593} (2001) 731]
  [hep-ph/0002108].
  %%CITATION = NUPHA,B579,17;%%

\bibitem{Freund:2001pn}
  M.~Freund,
  %{\it Analytic approximations for three neutrino oscillation parameters and
  %probabilities in matter,}
  Phys.\ Rev.\  D {\bf 64}, 053003 (2001)
  [hep-ph/0103300].
  %%CITATION = PHRVA,D64,053003;%%

\bibitem{Akhmedov:2004ny}
  E.~K.~Akhmedov, R.~Johansson, M.~Lindner, T.~Ohlsson and T.~Schwetz,
  %{\it Series expansions for three-flavor neutrino oscillation probabilities  in
  %matter,}
  JHEP {\bf 0404}, 078 (2004)
  [hep-ph/0402175].
  %%CITATION = JHEPA,0404,078;%%

\bibitem{Burguet-Castell:2001ez}
  J.~Burguet-Castell, M.~B.~Gavela, J.~J.~Gomez-Cadenas, P.~Hernandez and O.~Mena,
  %``On the measurement of leptonic CP violation,''
  Nucl.\ Phys.\  B {\bf 608} (2001) 301
  [hep-ph/0103258].
  %%CITATION = NUPHA,B608,301;%%

\bibitem{Huber:2002mx}
  P.~Huber, M.~Lindner and W.~Winter,
  %{\it Superbeams versus neutrino factories,}
  Nucl.\ Phys.\ B {\bf 645} (2002) 3
  [hep-ph/0204352].
  %%CITATION = HEP-PH 0204352;%%

\bibitem{globes}
%\bibitem{Huber:2004ka}
  P.~Huber, M.~Lindner and W.~Winter,
  %{\it Simulation of long-baseline neutrino oscillation experiments with
  %GLoBES,}
  Comput.\ Phys.\ Commun.\  {\bf 167}, 195 (2005)
  [hep-ph/0407333];
  %%CITATION = CPHCB,167,195;%%
%
%\bibitem{Huber:2007ji}
  P.~Huber, J.~Kopp, M.~Lindner, M.~Rolinec and W.~Winter,
  %{\it New features in the simulation of neutrino oscillation experiments with
  %GLoBES 3.0,}
  hep-ph/0701187.
  %%CITATION = HEP-PH/0701187;%%

\bibitem{T2K}
  Y.~Itow et al.,
  %{\it The JHF-Kamioka neutrino project,}
  hep-ex/0106019.
  %%CITATION = HEP-EX 0106019;%%

\bibitem{Ayres:2004js}
  D.~S.~Ayres et al.\ [NOvA Collaboration],
  %{\it NOvA proposal to build a 30-kiloton off-axis detector to study neutrino
  %oscillations in the Fermilab NuMI beamline,}
  hep-ex/0503053.
  %%CITATION = HEP-EX/0503053;%%

\bibitem{Donini:2004hu}
  A.~Donini, E.~Fernandez-Martinez, P.~Migliozzi, S.~Rigolin and L.~Scotto Lavina,
  %{\it Study of the eightfold degeneracy with a standard beta-beam and a
  %super-beam facility,}
  Nucl.\ Phys.\  B {\bf 710} (2005) 402
  [hep-ph/0406132].
  %%CITATION = NUPHA,B710,402;%%

\bibitem{Barger:2002xk}
  V.~Barger, D.~Marfatia and K.~Whisnant,
  %{\it How two neutrino superbeam experiments do better than one,}
  Phys.\ Lett.\  B {\bf 560} (2003) 75
  [hep-ph/0210428].
  %%CITATION = PHLTA,B560,75;%%

\bibitem{t2k+nova}
%\bibitem{Huber:2002rs}
  P.~Huber, M.~Lindner and W.~Winter,
  %{\it Synergies between the first-generation JHF-SK and NuMI superbeam
  %experiments,}
  Nucl.\ Phys.\  B {\bf 654} (2003) 3
  [hep-ph/0211300];
  %%CITATION = NUPHA,B654,3;%%
%
%\bibitem{Minakata:2003ca}
  H.~Minakata, H.~Nunokawa and S.~J.~Parke,
  %{\it The complementarity of eastern and western hemisphere long-baseline
  %neutrino oscillation experiments,}
  Phys.\ Rev.\  D {\bf 68} (2003) 013010
  [hep-ph/0301210];
  %%CITATION = PHRVA,D68,013010;%%
%
%\bibitem{Mena:2004sa}
  O.~Mena and S.~J.~Parke,
  %{\it Untangling CP violation and the mass hierarchy in long baseline
  %experiments,}
  Phys.\ Rev.\  D {\bf 70} (2004) 093011
  [hep-ph/0408070];
  %%CITATION = PHRVA,D70,093011;%%
%
%\bibitem{Mena:2006uw}
  O.~Mena, H.~Nunokawa and S.~J.~Parke,
  %{\it NOvA and T2K: The race for the neutrino mass hierarchy,}
  Phys.\ Rev.\  D {\bf 75} (2007) 033002
  [hep-ph/0609011].
  %%CITATION = PHRVA,D75,033002;%%

\bibitem{MenaRequejo:2005hn}
  O.~Mena-Requejo, S.~Palomares-Ruiz and S.~Pascoli,
  %{\it Super-NOvA: A long-baseline neutrino experiment with two off-axis
  %detectors,}
  Phys.\ Rev.\ D {\bf 72} (2005) 053002
  [hep-ph/0504015];
  %%CITATION = HEP-PH 0504015;%%
%
%\bibitem{Mena:2005ri}
  %O.~Mena, S.~Palomares-Ruiz and S.~Pascoli,
  %{\it Determining the neutrino mass hierarchy and CP violation in NOvA with a
  %second off-axis detector,}
  Phys.\ Rev.\ D {\bf 73} (2006) 073007
  [hep-ph/0510182].
  %%CITATION = HEP-PH 0510182;%%


\end{thebibliography}
\end{document}